# A NEW CONTENT FORMAT FOR IMMERSIVE EXPERIENCES

*Joan Llobera*
joan.llobera@i2cat.net

## ABSTRACT

The arrival of head-mounted displays (HMDs) to the consumer market requires a novel content format that is, first, adapted to the specificities of immersive displays and, second, that takes into account the current reality of multi-display media consumption. We review the requirements for such content format and report on existing initiatives, some of our own, towards implementing such content format.

## 1. INTRODUCTION

The arrival of head-mounted displays (HMDs) to the consumer market requires a novel content format that is, first, adapted to the specificities of immersive displays and, second, that takes into account the current reality of multi-display media consumption. Immersive displays impose radically different audience requirements compared to traditional broadcast TV and social media. For example, cuts between shots, which constitute the very basic fabric of traditional cinematic language, do not work well in immersive displays. Immersive displays also need to integrate with the consumer habits already in place within the contemporary living room, where TV consumers often use second screens -mostly smartphones, tablets or laptops. This paper summarizes, first, the requirements for such content format and, second, review current initiatives to deliver such content.

## 2. FORMAT REQUIREMENTS

### 2.1. Immersive display requirements

Presence in Immersive Virtual Environments (IVEs) is based upon two factors: *place illusion* and *plausibility* [1]. Place illusion is realized through the affordance of sensorimotor correlations. Research in this area has also shown the extreme importance of having a virtual body to feel place illusion.

Plausibility is a less understood cognitive factor that requires implementing interaction mechanisms similar to our everyday social conventions. Creating compelling immersive content requires addressing both these requirements.

### 2.2. Audio-visual conventions

One reason cuts and camera movements do not work well in IVEs is because they override sensorimotor affordances. One option is to avoid using these by creating content closer to the conventions of theatre. Otherwise, a suitable alternative to cuts and camera movements, preserving place illusion, needs to be introduced.

### 2.3. Multi-platform delivery

Immersive content needs to shareable online through social media. It also needs to be possible to visualize and edit it with common media consumption devices (TV, tablets and smartphones).

### 2.4 Interaction paradigms

Interaction with the overall IVEs and particularly with virtual characters needs to be intuitive. In addition, the implementation of interactive content needs to be done cost-effectively, close to standards in audiovisual and videogame industry.

## 3. EMERGING PRODUCTION TOOLS

### 3.1. Delivering place illusion

Traditionally, IVEs were created using three-dimensional (3D) Computer Graphics Imagery (CGI). 3D Modeling and character animation were combined with real time sensing data to allow sensorimotor affordances (turn your head, look under the table…). This remains a viable option to produce content.

More recently, omnidirectional video production tools (for example, www.video-stitch.com/) have delivered simpler methods to deliver photo-realistic content allowing for head rotations, shortcutting some of the technical burden associated with traditional CGI content production. Emerging commercial solutions, such as 8i (www.8i.com) and Presenz (www.nozon.com/), which build upon research in free viewpoint video and point cloud rendering, also allow for head displacements.

Despite these solutions only allows for head movements, simple methods and the peripherals of a new generation of HMDs (HTC vive, Oculus Touch, Microsoft Kinect…) allow to integrate easily animated virtual bodies linked to the movements of users immersed in such formats. Place illusion is therefore achievable in current consumer HMDs.

---
This work has been funded by European Union's Horizon 2020 program under grant agreement nº688619.

### 3.2 Replacing cuts

In terms of audiovisual conventions, one option is to introduce *portals*, i. e., inserts within a 3D scene that allow displaying either directional content, either another 3D scene, similar to how it was implemented in the classic videogame Portal, by Valve Corporation.

Such simple mechanism allows reintroducing traditional content, with cuts and camera movements, appearing like a floating panel in the main 3D scene. It also allows rendering other 3D scenes to introduce flashbacks, close-ups, another character's perspective, etc., while preserving sensorimotor affordances.

In other terms: the adoption of *portals* as a language convention seems to allow reintroducing in IVEs some of the traditional narrative tools of cinema and TV, while preserving place illusion. In the ImmersiaTV project (www.immersiatv.eu) we are currently customizing the entire broadcast production and distribution chain to fully implement this idea.

### 3.3 Content delivery across devices

A different problem is the existence of several devices to visualize the experience. Currently, the only option available to get this functionality is to cast a virtual camera from the HMD device to the other devices. This solution is far from optimal, since it requires real-time encoding in the HMD and the use of resources otherwise available to render the IVE, it only works locally, and is not compatible with social media. In ImmersiaTV we are currently implementing a different solution, based on delivering synchronized content across devices (see Figure 1), and building upon emerging broadcast standards (see reference DTS/JTC-DVB-343, from the European Telecommunications Standard Institute).

### 3.4 Delivering plausible interactions

Despite the undeniable appeal of using innovative video formats to provide immersive experiences, this production strategy also has its own limits. As mentioned earlier, the need to introduce a virtual body implies a hybrid video-CGI approach is necessary. However, this still does not address the fact that the eventual active engagement of the end-user needs to meet with a content format that integrates her input while preserving plausibility.

The natural way of interacting in an immersive virtual environment is through spontaneous social and body-centered interaction, and automating interactive conversations with smart virtual characters is still a cumbersome process, unfit for uptake in an industrial production context. However, there exist approaches that are robust, scalable and require little technical knowledge for content production [2,3]. Given the growing demand for virtual reality content, it seems now is the time to evolve them to fit within current industrial production tools.

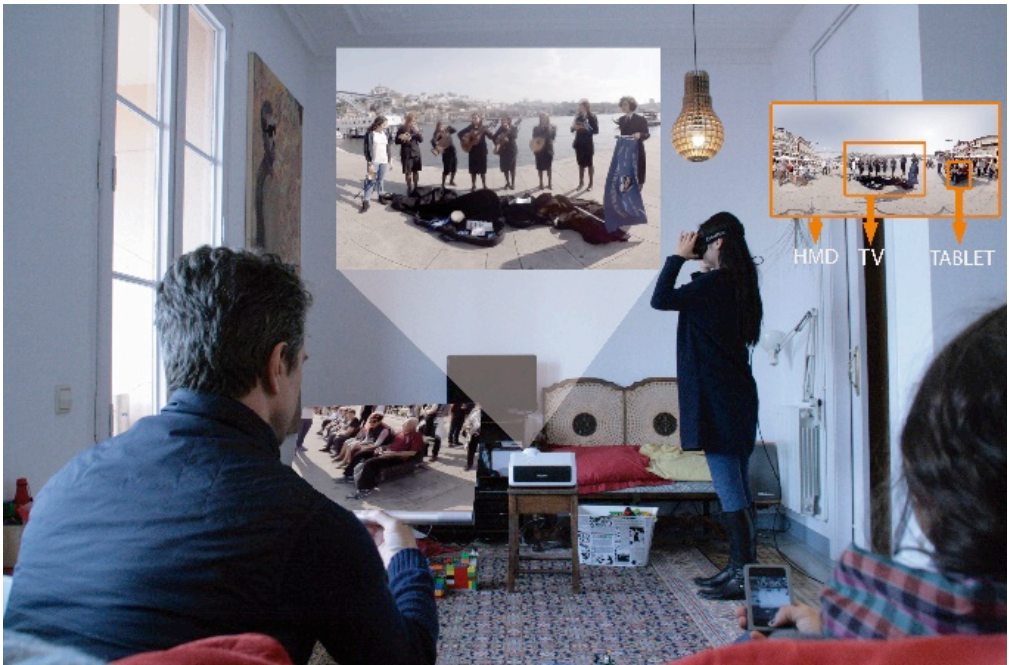

**Fig. 1.** A home environment with synchronized content across devices, as explored in ImmersiaTV